\documentstyle[preprint,aps,fleqn,amssymb]{revtex} 

\begin{document}
\draft

\title{Heat Capacity Evidence for the Suppression of Skyrmions 
at Large Zeeman Energy}
\author{S. Melinte$^1$, E. Grivei$^1$, V. Bayot$^1$ and M. Shayegan$^2$}
\address{$^1$Unit\'e PCPM, Universit\'e Catholique de Louvain, 1348
Louvain-la-Neuve, Belgium\\
$^2$Department of Electrical Engineering, Princeton University,
Princeton N.J. 08544\\}

\date{\today}
\maketitle

\begin{abstract}
Measurements on a multilayer two-dimensional electron system (2DES)
near Landau level filling $\nu$=1 reveal the disappearance of the nuclear
spin contribution to the heat capacity as the ratio $\tilde{g}$
between the Zeeman and Coulomb energies
exceeds a critical value $\tilde{g}_c$$\approx$0.04.
This disappearance suggests the vanishing of the Skyrmion-mediated
coupling between the lattice and the nuclear spins as the
spin excitations of the 2DES make a transition from
Skyrmions to single spin-flips above $\tilde{g}_c$.
Our experimental $\tilde{g}_c$ is smaller than the calculated
$\tilde{g}_c$=0.054 for an ideal 2DES; we discuss possible origins of
this discrepancy.
\end{abstract}



The ground state and spin excitations of a two-dimensional electron
system (2DES) near Landau level (LL) filling $\nu$=1 have attracted much recent
interest~\cite{Activation,Sondhi,NMR,Brey95,Cote97,Schmeller,Maude,Aifer,ThickSkyrme}.
At this filling, the Coulomb exchange energy plays a dominant role,
leading to a substantially larger quantum Hall effect (QHE)
excitation gap than the expected single-particle Zeeman
splitting~\cite{Activation}.
Moreover, the lowest energy charged excitations of the 2DES at $\nu$=1
are expected to be spin textures known as {\it
Skyrmions}~\cite{Sondhi}, provided that the Zeeman energy is small.
Pioneering nuclear magnetic resonance (NMR) experiments revealed
a pronounced reduction of the 2DES spin polarization and a strong
enhancement of the nuclear spin-lattice relaxation rate near
$\nu$=1~\cite{NMR}.
These observations have been attributed to the presence of Skyrmions
in the electronic ground state~\cite{Brey95}, and to
Skyrmion-induced low-energy spin-flip excitations~\cite{Cote97}, respectively.
Magnetotransport~\cite{Schmeller,Maude} and magnetooptical~\cite{Aifer}
data have provided further evidence for the existence of
finite-size Skyrmions.

Low temperature ($T$) heat capacity ($C$) measurements on a GaAs/AlGaAs
multiple-quantum-well (MQW) heterostructure revealed that $C$
near $\nu$=1 is dominated by the Schottky nuclear heat
capacity~\cite{Bayot} of Ga and As atoms in the quantum wells (QWs).
This observation implies a strong enhancement of the nuclear spin-lattice
relaxation rate near $\nu$=1, consistent with NMR experiments~\cite{NMR},
and suggests that heat capacity
is a sensitive probe of the presence or the absence of Skyrmions in 2DESs.

Hitherto, most of the experimental studies at $\nu$=1 concentrated on the
range of small ratio
$\tilde{g}$=$\mid$$g^\ast$$\mid$$\mu_{B}B/(e^{2}/\epsilon l_B)$
between the Zeeman and Coulomb energies~\cite{NMR,Maude,Aifer,Bayot},
where $\mid$$g^\ast$$\mid$$\mu_B$=0.3 K/T,
$\epsilon$$\approx$13$\epsilon_0$,
$l_B$=$(\hbar/eB_{\perp})^{1/2}$ is
the magnetic length, and $B$ and $B_{\perp}$ are the total and
perpendicular
components of the magnetic field, respectively.
One important theoretical prediction is that a transition
from Skyrmions to single spin-flip excitations is
expected above a critical $\tilde{g}_c$~\cite{Sondhi,ThickSkyrme}.
Up to now, the experiments performed at large $\tilde{g}$ probed
the QHE activation energy at $\nu$=1 ($\Delta$) which exhibits only a
smooth and gradual evolution with $\tilde{g}$~\cite{Schmeller}.
This fact prohibits an accurate determination
of $\tilde{g}_c$ and restricts the investigations to the case of
filling exactly equal to unity.
In this paper, we report evidence for $\tilde{g}_c$ in the
range 0.7$\lesssim$$\nu$$\lesssim$1.3
in a low-density, modulation-doped GaAs/AlGaAs MQW heterostructure.
The heat capacity near $\nu$=1 and $\Delta$
at $\nu$=1 were measured as the Zeeman
energy was tuned by tilting the sample in the magnetic field.
The data show that the nuclear contribution to the heat capacity
decreases in a narrow $\tilde{g}$-range and is suppressed for
$\tilde{g}$$\gtrsim$0.04. This result provides evidence for
a rather abrupt transition from Skyrmions to single spin-flip excitations at
$\tilde{g}_c$$\approx$0.04.
This $\tilde{g}_c$ is somewhat smaller than $\tilde{g}_c$=0.054
expected from theoretical calculations for an {\it ideal} 2D
system~\cite{Sondhi,ThickSkyrme}.
The discrepancy likely comes from corrections to $\tilde{g}_c$
because of finite layer-thickness, LL mixing, and disorder
in a {\it real} sample. We also discuss the possibility of a LL
crossing at high tilt angles ($\theta$) which might be
responsible for the suppression of Skyrmions in our sample for
$\tilde{g}$$\gtrsim$0.04. Finally, the heat capacity data at
intermediate $\theta$ (0.02$\lesssim$$\tilde{g}$$\lesssim$0.03) exhibits
some anomalous features which we presently do not understand.


The heterostructure used in this study consists
of one hundred 300 $\rm \AA$-thick GaAs QWs, separated by 2500 $\rm \AA$-thick
$\rm Al_{0.1}Ga_{0.9}As$
barriers which are $\delta$-doped with donors (Si) near their centers.
Heat capacity experiments were carried on a 7$\times$10 $\rm mm^2$ piece
of the wafer thinned to 160 $\mu$m. The thermometer and  heater
are carbon paint resistors, deposited on the substrate
side of the sample and electrically connected with
graphite fiber bundles which also serve as thermal link to the heat
sink.
In the restricted $T$ and $\nu$-range investigated here, we used a
corrected relaxation method~\cite{Bayot,Grivei} for $C$-experiments: $C =
\kappa
\tau_{ext}$, where $\kappa$ and $\tau_{ext}$ are the measured thermal
conductance to the heat sink and the external time constant,
respectively.
It is important to note that we measure the heat
capacity of those nuclei with a nuclear spin-lattice relaxation time
$T_1$$<$$\tau_{ext}$.
Typical $\tau_{ext}$ in our experiments was $\sim$100 to 1000 s.
We also performed electrical measurements on a 2$\times$2 $\rm mm^2$
piece from the same wafer.
Both measurements were performed in a dilution refrigerator and the
samples were tilted {\it in situ} so that an
angle $0^{\circ}$$\lesssim$$\theta$$\lesssim$$77^{\circ}$ forms
between $B$ and the normal to the sample plane.


Figure 1 shows the longitudinal resistance  ($R_{xx}$)  as a function of
$B$ at $\theta$=$0^{\circ}$. The density determined from the
position of $R_{xx}$ minima is  8.5$\times$$10^{10}$ $\rm cm^{-2}$
per layer and the estimated mobility is $\approx 7 \times$$10^{5}$ $\rm
cm^{2}/Vs$.
The very high quality of this heterostructure is evidenced by the presence
of strong fractional QHE states at $\nu$=2/5 and 3/5.
The value of the excitation gap ($\Delta$) was determined from the
$T$-dependence of $R_{xx}$ at $\nu$=1 in the thermally activated
regime where $R_{xx}\propto\rm exp(-\Delta$/2$T$).
At $\theta$=$0^{\circ}$, $\Delta$=20 K,
comparable to the measured $\Delta$ in high-quality conventional
single-layer 2DESs~\cite{Schmeller}.
The evolution of $\Delta$ with $\theta$ is presented in Fig.
2 by plotting $\Delta$ vs Zeeman energy, both expressed in units of
$e^{2}/\epsilon l_B$.
The inset to Fig. 2 shows the
calculated Hartree-Fock (HF) energy gap for the creation of a widely
separated Skyrmion/Antiskyrmion pair as a function of $\tilde{g}$ for
our sample~\cite{CooperPrivate}.
This calculation takes into account the finite thickness of the electron
layers; the relevant parameter is $w/l_B$=0.52 where $w$=71 $\rm \AA$ is
the rms width of the self-consistently calculated subband wavefunction
in each QW.

Similar to previous results for single QWs~\cite{Schmeller}, there is an overall qualitative
agreement between the measured and calculated gaps in Fig. 2~\cite{gap}.
In particular, assuming that the slope
$K$=$\partial \Delta / \partial $($\mid$$g^\ast$$\mid$$\mu_B B$) gives the
number of flipped electron spins within a 
single, charged excitation at $\nu$=1, both theory and experiment give $K$$\cong$9 for
$\tilde{g}$$\approx$0.012 and $K$$\cong$1 in the limit of large
$\tilde{g}$~\cite{dev}.
However, given the experimental uncertainty in the measured gaps, the
absence of quantitative agreement with theoretical predictions,
and the fact that $\Delta$ is expected to slowly approach the
single spin-flip dependence ($K$=1),
prohibit an accurate determination of $\tilde{g}_c$ based on transport
measurements.

We now present heat capacity experiments,
which reveal a dramatic and rather abrupt dependence on $\tilde{g}$
and provide evidence for the disappearance of Skyrmions.
In the investigated range of $T$ and $B$,
the results can be understood by using the Schottky nuclear heat capacity
model developed in Ref.~\cite{Bayot}. The model assumes  a
strong coupling of the nuclear spin system in the QWs to the lattice
near $\nu$=1 provided by low-energy spin excitations~\cite{NMR,Cote97}.
In the present sample, the Schottky nuclear heat capacity of Ga
and As atoms in the QWs is estimated at: $C_{QW}$=3.3$\times$$10^{-11}$
$B^{2} T^{-2}$ (J/K)~\cite{Bayot}. Therefore, the ratio ($\xi$) between
the measured $C$ and calculated $C_{QW}$ provides the fraction of QWs nuclei
which strongly couple to the lattice, and hence signals the presence
of low-energy spin excitations in the 2DES attributed to
Skyrmions~\cite{Sondhi,Cote97}.

Figures 3 and 4 capture the evolution of the heat capacity,
represented by the parameter $\xi$$\equiv$$C/C_{QW}$, with
tilt angle, at $T$=60 mK.
At $\theta$=$0^{\circ}$ (Fig. 3), the data are qualitatively similar to those
reported for a 100-period  GaAs/$\rm Al_{0.3}Ga_{0.7}As$
heterostructure with a density of 1.4$\times$$10^{11}$
$\rm cm^{-2}$ per layer~\cite{Bayot,HighD}.
The decreasing density of Skyrmions as $\nu$$\rightarrow$1 is responsible
for the minimum observed in $\xi$ at $\nu$$\approx$1~\cite{Sondhi}.
The non-vanishing $\xi$ at $\nu$=1 could arise from density
inhomogeneities in such a large, thinned MQW sample.
Far away from $\nu$=1 ($\nu$$\lesssim$0.7 and  $\nu$$\gtrsim$1.4), $\xi$ is
essentially zero as the 2DES is at fillings where Skyrmions are no longer
relevant~\cite{Sondhi,Bayot}.
We note that  $\xi$ at maxima reaches supraunitary values: $\xi$$\approx$1.3.
Besides the
experimental accuracy ($\pm$15\%) and uncertainty in well-width ($\pm$10\%),
the barriers' nuclei may enhance the
measured $C$ due to the spread of the electron wave function from
the QWs into the $\rm Al_{0.1}Ga_{0.9}As$ barriers which are only
about 0.1 eV high.

As seen in Fig. 3, $\xi$ vs $\nu$ at
$\theta$=$46^{\circ}$, is nearly identical to the $\theta$=$0^{\circ}$ data.
On the other hand, at $\theta$=$71^{\circ}$, the data show a significant
asymmetry with
respect to the $\nu$=1 position and a broadening of the $\nu$$>$1 peak.
For $\nu<$1, $\xi$ at $\nu$$\approx$0.9 is reduced by a factor of 2
when compared to the $\theta$=$0^{\circ}$ value and  vanishes for
$\nu$$\lesssim$0.8.
Most remarkable, however, is that the magnitude of $\xi$ at the
$\nu$$>$1 peak is comparable to the $\theta$=$0^{\circ}$ data,
implying a still strong coupling of the nuclei to the lattice.
This is a particularly noteworthy observation as it highlights that the
heat capacity is a very sensitive probe of the low-energy
spin excitations, and therefore Skyrmions, in a regime where the
transport data and calculations both reveal a small Skyrmion size
($K$$<$3) and a very weak dependence of $\Delta$ on
$\tilde{g}$ (see Fig. 2 and its inset near $\tilde{g}$$\sim$0.035
($\theta$=$71^{\circ}$)).

When $\theta$ is further increased above $71^{\circ}$ only by few
degrees (Fig. 4),
the nuclear heat capacity decreases dramatically for all $\nu$.
For $\theta>$$74^{\circ}$, the
nuclear heat capacity is no longer measurable up to the highest
investigated tilt-angle ($77^{\circ}$).
To bring into focus the evolution of the coupling between the nuclear
spin system and the lattice with
$\theta$ and $\tilde{g}$, we plot $\xi$  at $\nu>$1 and $\nu<$1 maxima vs
$\tilde{g}$ (Fig. 5).
The coupling due to low-energy spin
excitations is progressively suppressed for $\tilde{g}\gtrsim$0.035 and
vanishes in the range 0.037$\lesssim$$\tilde{g}$$\lesssim$0.043.
We believe that this behavior
provides evidence for the {\it transition from Skyrmions to single
spin-flip excitations} at $\tilde{g}_c$$\approx$$0.04$ in our sample.
This $\tilde{g}_c$ is smaller than the theoretical $\tilde{g}_c$=0.054
calculated for the Skyrmion to single spin-flip transition for an
{\it ideal} 2DES~\cite{Sondhi,ThickSkyrme}.
Several factors, however, are expected to reduce $\tilde{g}_c$ for a
{\it real} 2DES. This includes the finite thickness of the electron
layer~\cite{ThickSkyrme}, LL mixing~\cite{Bonesteel}, LL crossing~\cite{Tomas}, and
disorder. Indeed, calculations by
Cooper~\cite{ThickSkyrme,CooperPrivate} (also see inset to Fig. 2)
reveal that taking into account the finite $z$-extent of the 2DES {\it
alone} leads to $\tilde{g}_c$=0.047, closer to our experimental value.
It is worth emphasizing, however, that all these calculations are
performed at $\nu$=1 while the heat capacity data of Figs. 4 and 5
provide values for $\tilde{g}_c$ in the full $\nu$-range where Skyrmions
are
the expected ground state of the 2DES. In particular, we observe that
$\tilde{g}_c$ depends on filling factor and increases from
$\tilde{g}_c$$\approx$0.037 at $\nu$=1.2 to
$\tilde{g}_c$$\approx$0.043 at $\nu$=0.9.

Finally, we wish to report an anomalous and unexpected behavior for
the measured heat capacity at intermediate tilt-angles
($50^{\circ}$$\lesssim$$\theta$$\lesssim$$66^{\circ}$) which shows evidence for the
complex dependence of Skyrmions on $\theta$ and $\tilde{g}$.
As depicted in Fig. 6, we observe a reduction of $\xi$ in a narrow $\nu$-range
compared to the $\theta$=$0^{\circ}$ data.
The vertical arrows in Fig. 6 point
to the total magnetic field at which $\xi$ is most significantly reduced.
The anomaly moves to  higher $\nu$ with
small increases in $\theta$  (from $\nu$=0.81 at $\theta$=$50^{\circ}$ to
$\nu$$\approx$1 at $\theta$=$66^{\circ}$) and shows an increasing intensity
when compared to the
$\theta$=$0^{\circ}$ data-envelope, e.g., at $\theta$=$50^{\circ}$, $\xi$
at $\nu$=0.81 is reduced by a factor of
$\approx$3, while at $\theta$=$66^{\circ}$, $\xi$ at $\nu$$\approx$1 is
essentially zero.
The observed reduction of $\xi$ shows that low-energy spin excitations
are strongly
affected or even suppressed, which would signal the disappearance of
Skyrmions for limited $\nu$ or $B$-ranges dependent on tilt-angle.
Even though the exact behavior of the anomaly might be
specific to our heterostructure, it reveals the
subtle influence of $\theta$ on spin excitations of 2DESs whose
description will require further theoretical and experimental work.
However, we note that the anomaly strongly affects the heat capacity
at $\nu$=1  in the range $61^{\circ}$$\lesssim$$\theta$$\lesssim 66^{\circ}$
which corresponds to the range of $\theta$ where the
activation energy at $\nu$=1 exhibits the strongest departure
from the HF calculations~\cite{dev}.
The suppression of Skyrmions at $\nu$=1 in this range of $\theta$
would certainly affect $\Delta$, and possibly increase it
up to the single spin-flip value (dashed line in Fig. 2).
This might explain the somewhat anomalous behavior of
$\Delta$ vs $\tilde{g}$ in the range 0.02$\lesssim$$\tilde{g}$$\lesssim$0.03.


In conclusion, the heat capacity experiments reveal the subtle and critical
influence of tilted magnetic fields on the ground and excited states of 2DESs
near $\nu$=1 in GaAs/AlGaAs heterostructures.
The data indicate the disappearance of low-energy spin
excitations in the 2DES, which in turn provides evidence for the
suppression of Skyrmions above a critical Zeeman energy ($\tilde{g}_c$$\approx$0.04
in our sample).


The authors are much indebted to  N.R. Cooper, T. Jungwirth and S.P. Shukla
for numerical calculations.
This work has been supported by NATO grant CRG 950328 and the NSF
MRSEC grant DMR-9400362.
V.B. acknowledge financial support by the Belgian National
Fund for Scientific Research.
The work performed in Louvain-la-Neuve
was carried out under financial support of the programme "PAI"
sponsored by the "Communaut\'{e} Fran\c{c}aise de Belgique".


{\it Note Added.} - Since the submission of this Letter we performed NMR
experiments on the same heterostructure~\cite{Melinte}.
The measurements indicate that Skyrmions are suppressed  at $\nu$=1 
($70^{\circ}$$\lesssim$$\theta$$\lesssim$$73^{\circ}$) above 
$\tilde{g}_c$$\approx$0.038 which is totally consistent with the 
present heat capacity data. Moreover, the NMR results suggest that the LL crossing
occurs at significantly larger tilt angles ($\theta$$>$$80^{\circ}$) and hence
does not affect the observed $\tilde{g}_c$.



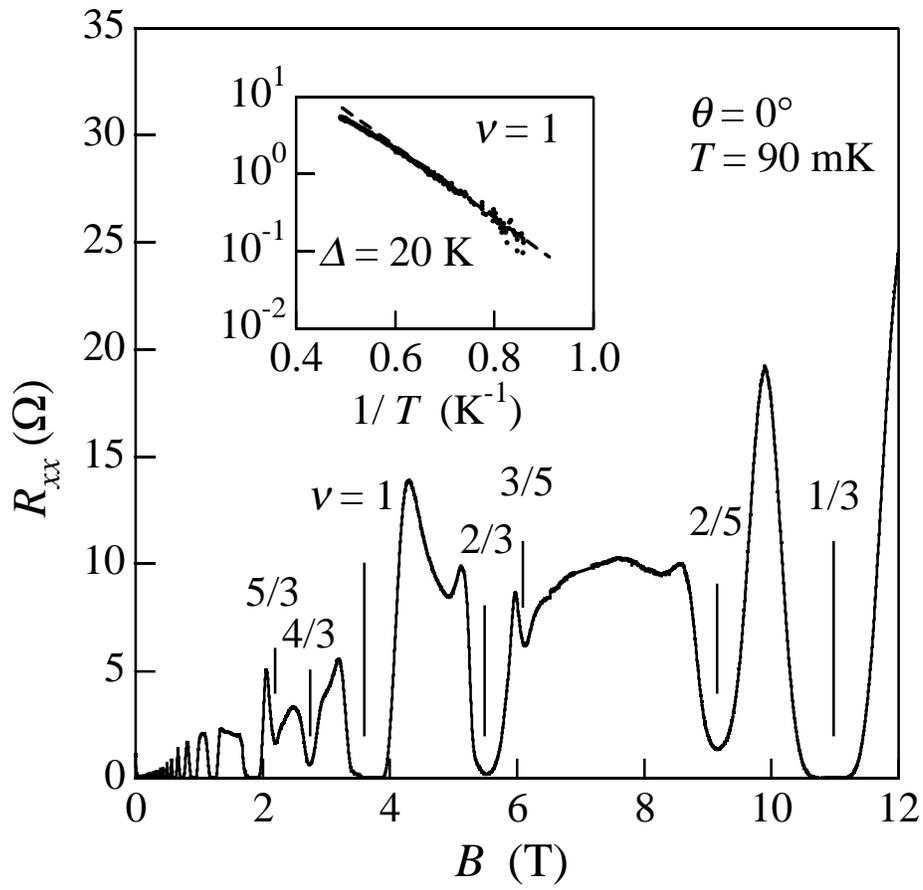
\begin{figure}
\caption{$R_{xx}$ vs $B$ at $T$=90 mK and $\theta$=$0^{\circ}$. The vertical
lines indicate some of the filling factors at which the QHE is observed.
The inset shows the temperature dependence of $R_{xx}$ used to determine
$\Delta$. From the slope of the least-square fit (dashed line),
a gap of 20 K is obtained.}
\label{transport}
\end{figure}

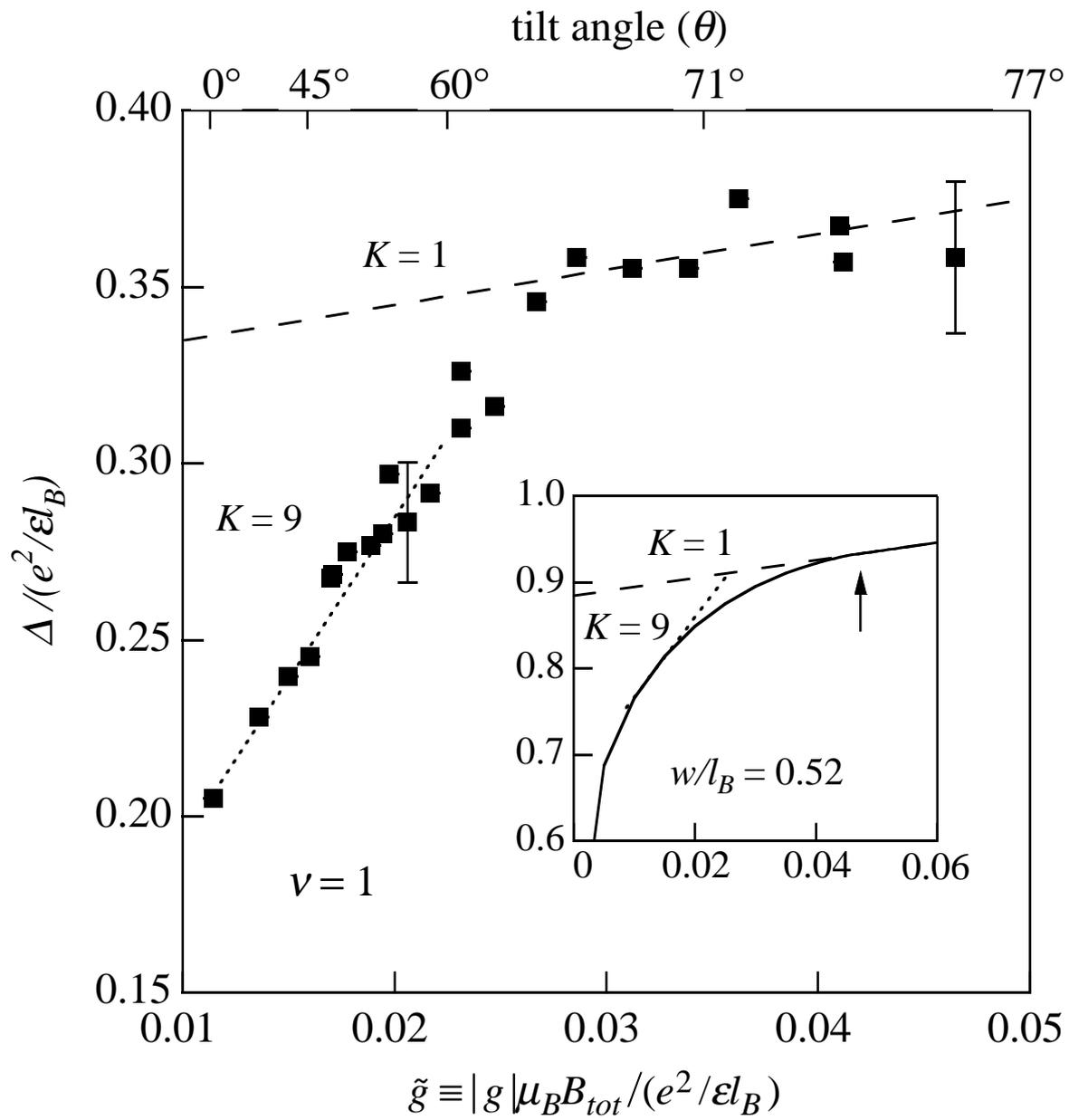
\begin{figure}
\caption{$\Delta$ vs Zeeman energy, both in units of $e^{2}/\epsilon l_{B}$.
The corresponding tilt angles are indicated on the top axis.
Inset: The energy gap to create a Skyrmion/Antiskyrmion pair (in units of
$e^{2}/\epsilon l_{B}$) as a function of $\tilde{g}$, from HF calculations
with $w/l_B$=0.52~\protect \cite{CooperPrivate}.
The vertical arrow indicates  $\tilde{g}_c$$\approx$0.047 above which
a transition from Skyrmions to single spin-flip excitations is predicted.
The dotted and dashed lines in the main figure and in the inset correspond
to $K$=9 and $K$=1, respectively.}
\label{activation}
\end{figure}

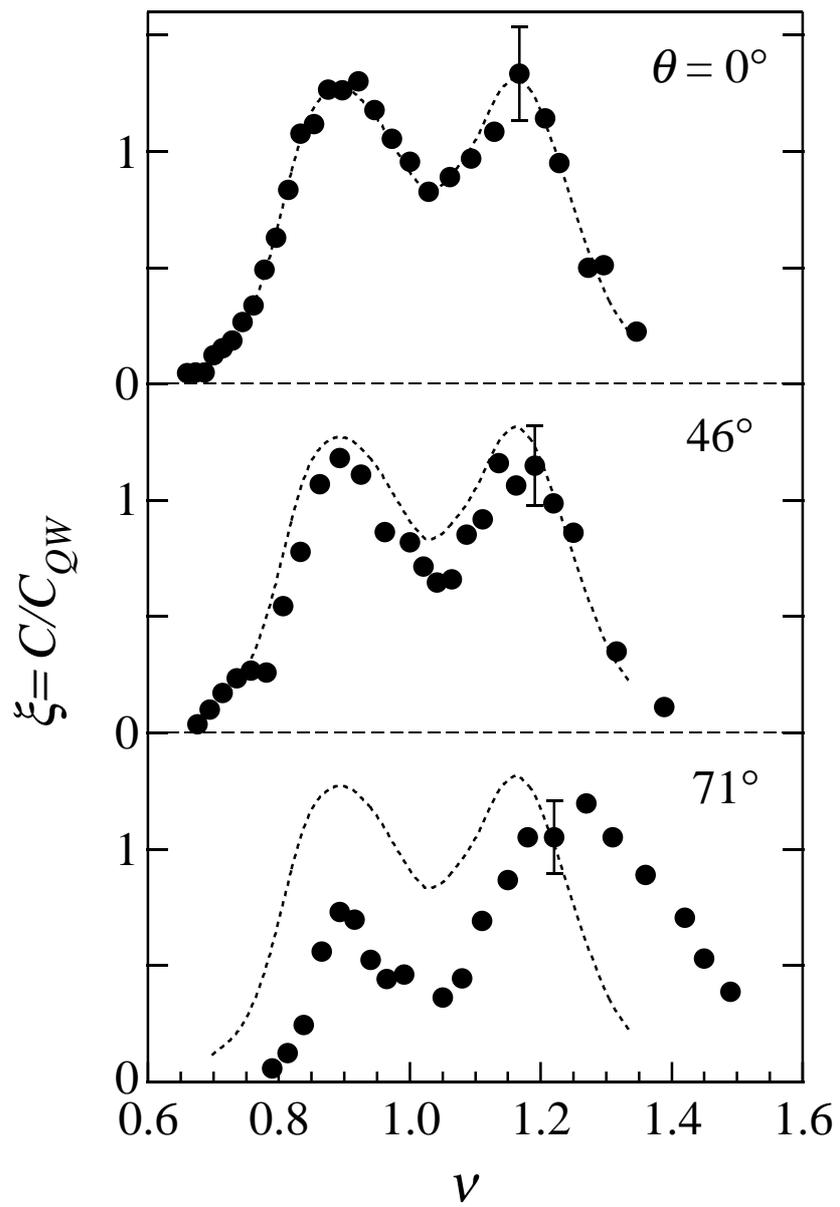
\begin{figure}
\caption{$\xi$=$C/C_{QW}$ vs $\nu$ at $T$=60 mK at the indicated
tilt angles.
The absolute accuracy for $\xi$ ($\pm$15\%) is shown. The
$\xi$ envelope for $\theta$=$0^{\circ}$ (dotted line), is reproduced
for comparison.}
\label{evolution}
\end{figure}

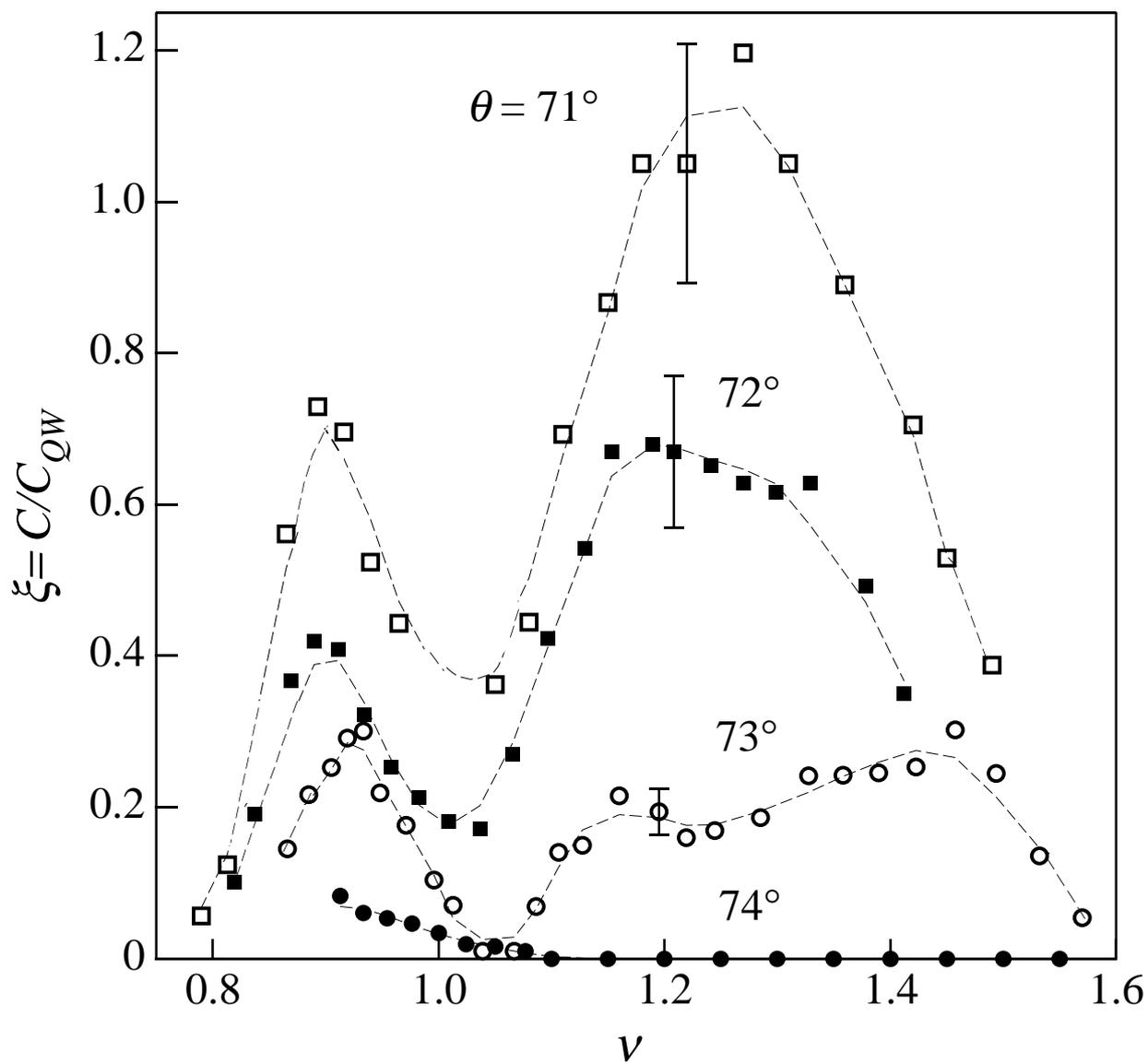
\begin{figure}
\caption{$\xi$  vs $\nu$
at $T$=60 mK for $71^{\circ}$$\leq$$\theta$$\leq$$74^{\circ}$.
The dashed lines are guides to the eye.}
\label{evolution2}
\end{figure}

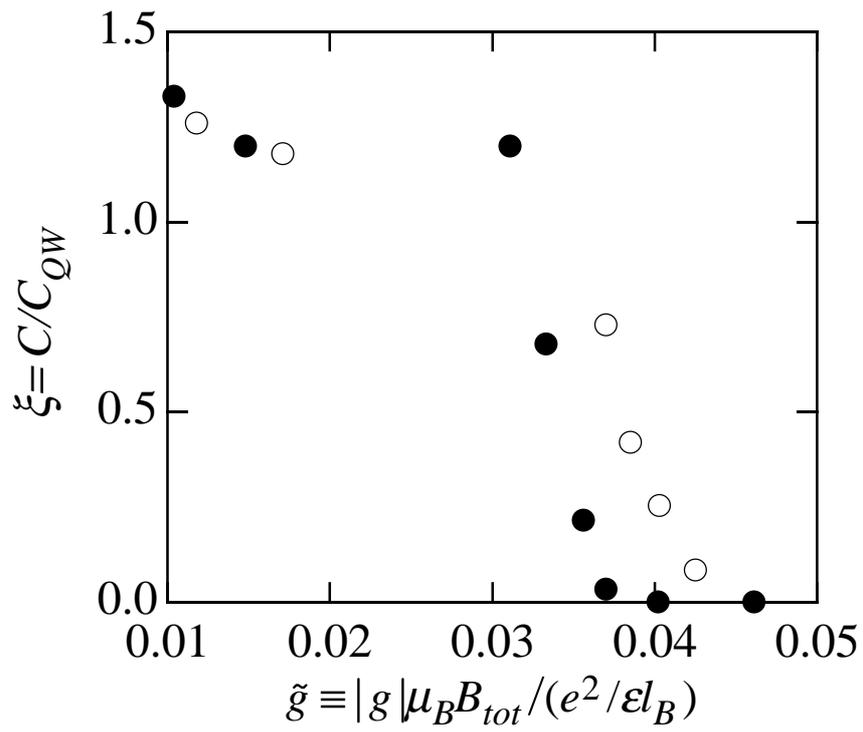
\begin{figure}
\caption{$\xi$  at $\nu>$1 ($\bullet$) and $\nu<$1 ($\circ$) maxima is
plotted vs $\tilde{g}$ at
$\theta$=$0^{\circ}$, $\theta$=$46^{\circ}$ and $\theta$$\geq$$71^{\circ}$.}
\label{evolution3}
\end{figure}

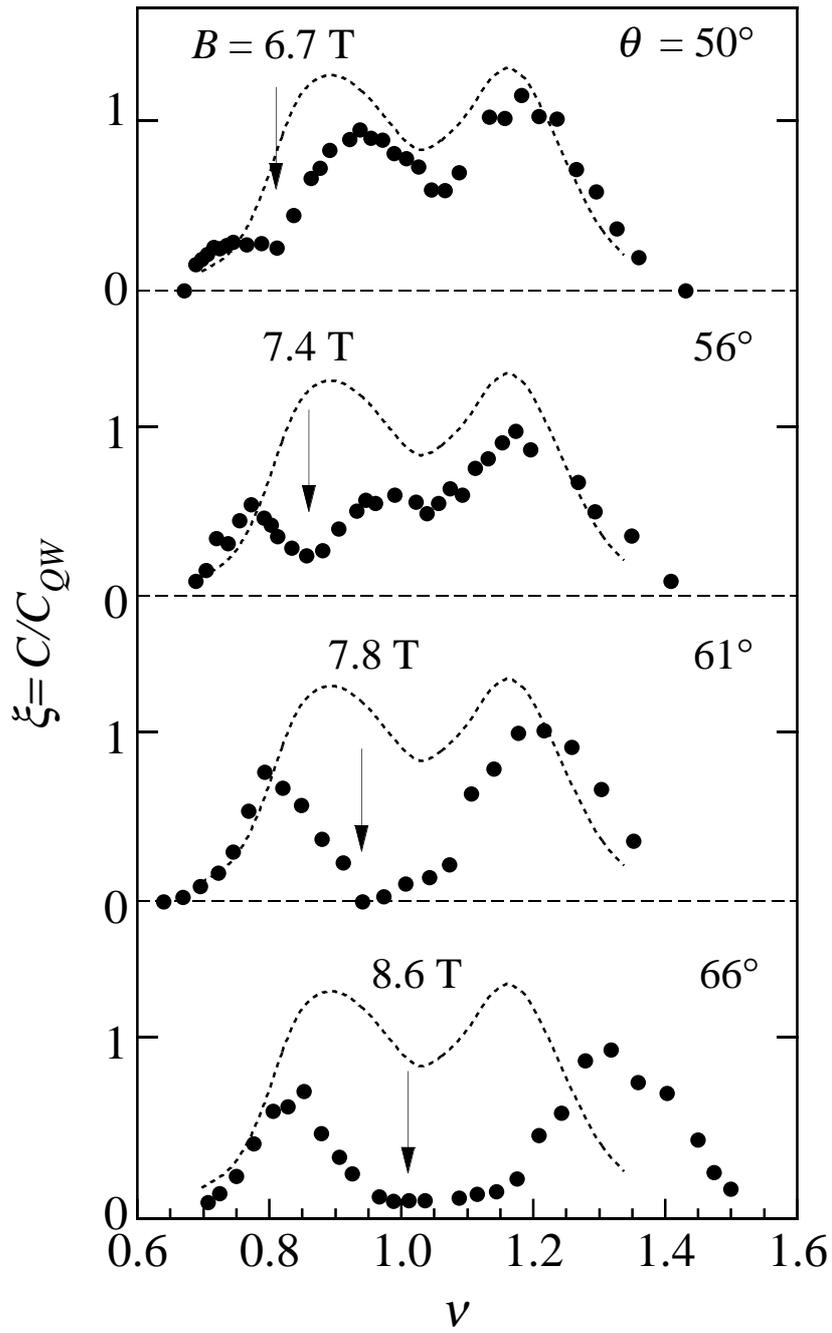
\begin{figure}
\caption{$\xi$=$C/C_{QW}$ vs $\nu$ at $T$=60 mK at the indicated
tilt angles.
The $\xi$ envelope for $\theta$=$0^{\circ}$ (dotted line), is reproduced
for comparison.}
\label{evolution4}
\end{figure}


\begin{thebibliography}{10}
\bibitem{dev} Note in Fig. 2 that near $\tilde{g}$=0.025
($61^{\circ}$$<$$\theta$$<$$66^{\circ}$), the experimental gap
appears to deviate from the smooth and gradual dependence expected
from the theory (inset to Fig. 2).
\end{thebibliography}
\end{document}